\documentclass[prl,twocolumn,superscriptaddress,floatfix,nopacs,preprintnumbers,footinbib]{revtex4-2}

\usepackage[utf8]{inputenc}
\usepackage{subfigure}
\usepackage[normalem]{ulem}
\usepackage[dvipsnames]{xcolor}
\usepackage{mathrsfs}
\usepackage{amssymb,bm}
\usepackage{amsmath}
\usepackage{mathtools}
\usepackage{slashed}
\usepackage{xfrac}
\usepackage{graphics}
\usepackage{graphicx}
\usepackage{physics}
\usepackage[h]{esvect}
\usepackage{accents}

\definecolor{lcolor}{rgb}{0.5,0,0}
\definecolor{citcolor}{rgb}{0,0.3,0.0}

\renewcommand{\tr}[1]{\mathrm{Tr}\left[#1\right]}

\usepackage[breaklinks,colorlinks,citecolor=citcolor,urlcolor=blue,linkcolor=lcolor]{hyperref}

\begin{document}

\author{Dana Avramescu}
\email[Corresponding author: ]{dana.d.avramescu@jyu.fi}
\affiliation{Department of Physics, University of Jyväskylä,  P.O. Box 35, 40014 University of Jyväskylä, Finland}
\affiliation{Helsinki Institute of Physics, P.O. Box 64, 00014 University of Helsinki, Finland}

\author{Vincenzo Greco}
\email{greco@lns.infn.it}
\affiliation{Department of Physics and Astronomy, University of Catania, Via Santa Sofia 64, I-95123 Catania, Italy}
\affiliation{INFN-Laboratori Nazionali del Sud, Via Santa Sofia 62, I-95123 Catania, Italy}

\author{Tuomas Lappi}
\email{tuomas.v.v.lappi@jyu.fi}
\affiliation{Department of Physics, University of Jyväskylä,  P.O. Box 35, 40014 University of Jyväskylä, Finland}
\affiliation{Helsinki Institute of Physics, P.O. Box 64, 00014 University of Helsinki, Finland}

\author{Heikki Mäntysaari}
\email{heikki.mantysaari@jyu.fi}
\affiliation{Department of Physics, University of Jyväskylä,  P.O. Box 35, 40014 University of Jyväskylä, Finland}
\affiliation{Helsinki Institute of Physics, P.O. Box 64, 00014 University of Helsinki, Finland}

\author{David M\"{u}ller}
\email{dmueller@hep.itp.tuwien.ac.at}
\affiliation{Institute for Theoretical Physics, TU Wien, Wiedner Hauptstraße 8, A-1040 Vienna, Austria}

\title{Heavy flavor angular correlations as a direct probe of the glasma}

\begin{abstract}
We use classical equations of motion for heavy quarks to show that the pre-equilibrium glasma phase of a heavy ion collision has an extremely strong effect on heavy quark angular correlations. At the same time the effect on the single inclusive spectrum is much more moderate. This suggests that $D\overline{D}$ meson angular correlations in future LHC measurements could provide a direct experimental access to the physics of the pre-equilibrium stage. 
\end{abstract}

\maketitle 

\textit{Introduction---}Ultrarelativistic heavy ion collisions are multi-stage processes where the Quark Gluon Plasma (QGP) state of (almost) thermalized matter is produced from a highly nonequilibrium initial state. In order to understand QGP production, different stages of the collision process need to be described using appropriate degrees of freedom. The QGP evolution itself can be understood in terms of relativistic fluid dynamics, and kinetic theory offers a convenient framework to describe the thermalization process~\cite{Kurkela:2015qoa}. Immediately after the collision and before the kinetic theory description in  terms of quasiparticles is applicable, very strong non-Abelian QCD color fields are present in the so-called pre-equilibrium glasma phase, which is a state of highly over-occupied gluonic matter~\cite{Kovner:1995ja,Lappi:2006fp,Fukushima:2011nq,Gelis:2012ri}. 

Typically, QGP properties have been extracted from detailed analyses of e.g.~collective flow measurements~\cite{Bernhard:2019bmu} that are dominated by light partons formed so late that they are not especially sensitive to the glasma. 
Generically accessing QCD dynamics in the glasma stage is extremely challenging because, in a rapidly expanding system, the strong glasma fields have a very short lifetime. In order to reveal direct unambiguous signatures of the glasma phase, it is imperative to look for processes that are sensitive to the very early-time dynamics. A unique possibility to look for such signals is provided by heavy quark production, as the large quark mass ensures that the quarks are formed early in the collision process and have time to interact with the glasma. 

The transport of hard probes during the pre-equilibrium phase has been investigated in numerous recent studies. Both heavy quark~\cite{Ruggieri:2018rzi,Sun:2019fud,Khowal:2021zoo} and jet~\cite{Ipp:2020mjc,Ipp:2020nfu} dynamics were studied in the glasma using classical transport equations solved numerically~\cite{Avramescu:2023qvv} or using analytical approximations~\cite{Carrington:2020sww,Carrington:2021dvw}. 
The large transport coefficients obtained for hard probes in the glasma were supported by calculations using effective kinetic theory \cite{Boguslavski:2023alu,Boguslavski:2023fdm,Boguslavski:2023waw}. Energy loss for jets during the glasma stage was considered in \cite{Barata:2024xwy}. The effect of energy loss during pre-equilibrium was previously addressed in~\cite{Andres:2019eus,Andres:2022bql}.
As transport coefficients are not directly measurable, it is crucial to quantitatively determine the phenomenological impact of glasma on actual observables sensitive to early time dynamics. 

Heavy quark angular  correlations have previously been discussed in the context of the equilibrium plasma phase \cite{Nahrgang:2013saa,Zhao:2024oma}. In this Letter, we demonstrate that these correlations, reflected experimentally in  $D\overline D$ meson pairs, are highly sensitive to the glasma phase. 
 On the other hand, the inclusive heavy quark spectrum is only weakly modified by  the glasma. This is especially intriguing as $D\overline D$ correlations have been measured in proton-proton and proton-lead collisions at the LHC~\cite{ALICE:2019oyn,CMS:2011yuk,LHCb:2012aiv}, and similar measurements in heavy ion collisions become possible for the first time at the high-luminosity phase LHC (HL-LHC)~\cite{ALICE:2022wwr}. Such measurements have so far been possible only indirectly via dimuon correlations~\cite{ATLAS:2023vms}.

\begin{figure*}[tb]
\includegraphics[width=0.95\textwidth]{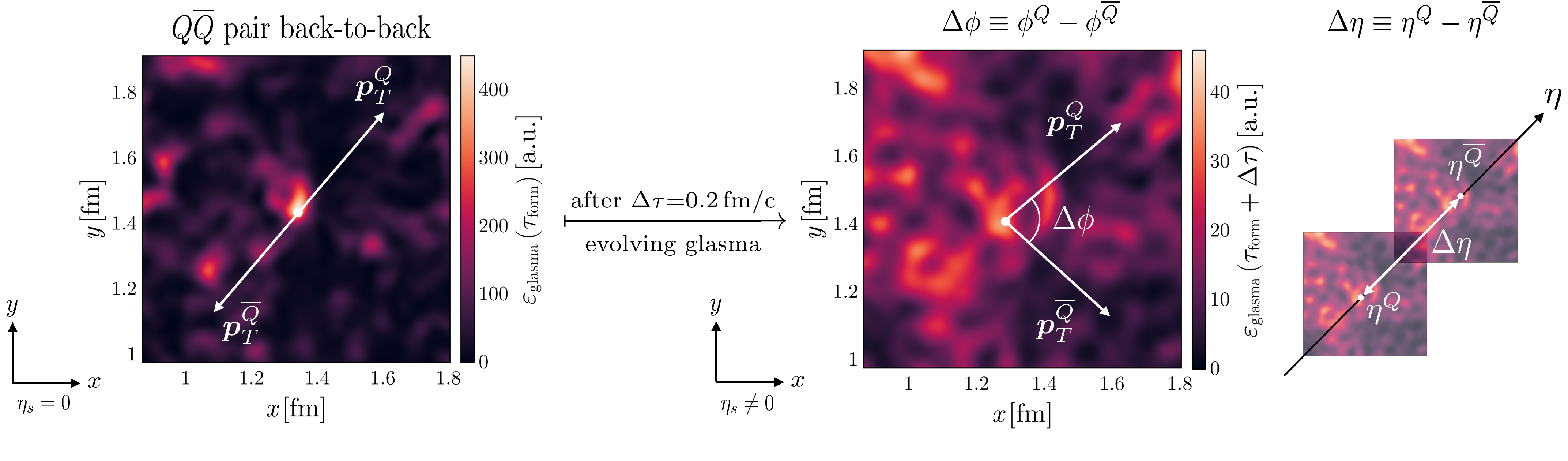}
\caption{Illustration of the time evolution of the initially back-to-back produced heavy quark pair $Q\overline Q$ in the evolving boost-invariant glasma background field. The evolution of the glasma energy density can be seen from the scale of the energy density (units are arbitrary but the same in both plots). 
\label{fig:sketch_quark_pair} }
\end{figure*}

This Letter is organized as follows. First, we discuss the glasma state and the numerical simulation that tracks the time evolution of heavy quarks in the evolving glasma background. Then, we quantify the effect of the glasma stage on the nuclear modification factor $R_{AA}$. This effect is found to be modest. However,  the effect on the heavy quark pair azimuthal correlations is found to be highly sensitive to glasma, leading us to argue that it is imperative to take the glasma phase into account for understanding future correlation measurements. Further technical details and more thorough studies are presented in a companion paper~\cite{Avramescu:2024poa}.

\textit{Glasma fields and test particles---}Hadronic collisions at high center-of-mass energy are sensitive to hadron structure at small $x$ (fraction of the nucleon momentum carried by the parton). When the parton density becomes of the same order as the inverse of the QCD coupling, it becomes necessary to take into account non-linear QCD dynamics. Furthermore, as the occupation number of soft gluons is very high, they can be treated classically. This is realized in the Color Glass Condensate effective  theory of QCD~\cite{Iancu:2003xm}. In this picture, the high-momentum modes (e.g.~valence quarks) move along the light cone and act as a source  $J^\mu$ of small-$x$ gluons described by the color field $A^\mu$. The time evolution of the system is described in terms of the classical Yang-Mills equations as
\begin{equation}
    \label{eq:cym}
    \mathcal{D}_\mu F^{\mu\nu}=J^\nu,
\end{equation}
where $\mathcal{D}_\mu=\partial_\mu-\mathrm{i}g[A_\mu,\,\cdot\,]$ is the covariant derivative, $F^{\mu\nu}=\partial_\mu A_\nu-\partial_\nu A_\mu -\mathrm{i}g[A_\mu,A_\nu]$ the field strength tensor and $g$ the coupling constant. 

We assume the classical color charge distribution to be Gaussian, and take it from the McLerran-Venugopalan model~\cite{McLerran:1993ni, McLerran:1993ka, McLerran:1994vd} valid for large nuclei. In this work, we consider infinitely large nuclei neglecting the impact parameter dependence. This allows us to estimate the glasma stage effect in central heavy ion collisions where, during the early stages, the partons do not have enough time to propagate to the nuclear periphery.

The Yang-Mills equations can be solved analytically at proper time  $\tau\to0^+$ to obtain the classical color fields immediately after the collision. The time evolution of the fields is obtained by solving the Yang-Mills equations numerically, using a real-time lattice gauge theory formulation of these equations~\cite{Krasnitz:1998ns,Lappi:2003bi} that ensure a gauge-independent discretization. 

Heavy quarks with position $x$, momentum $p$, and color charge $Q$ are treated as test particles whose time evolution on the evolving glasma background is described by Wong's equations~\cite{Wong:1970fu}
\begin{subequations}
    \label{eq:wongcurv}
    \begin{align}
        &\frac{\mathrm{d}x^\mu}{\mathrm{d}\tau}=\frac{p^\mu}{p^\tau}, \label{eq:wongcoord}\\
        &\frac{\mathrm{D}p^\mu}{\mathrm{d}\tau}=\frac{g}{T_R}\tr{QF^{\mu\nu}}\frac{p_\nu}{p^\tau}, \label{eq:wongmom}\\ 
        &\frac{\mathrm{d}Q}{\mathrm{d}\tau}=-\mathrm{i}g[A_\mu,Q]\frac{p^\mu}{p^\tau}.\label{eq:wongcharge}
    \end{align}
\end{subequations}
Wong's equations represent the equations of motion for classical test particles \cite{Heinz:1983nx,Kelly:1994dh,Litim:1999ns}. Thus there is no 
energy loss mechanism included in this formalism. 
We solve these equations  numerically using the 
leapfrog method~\cite{Krasnitz:1998ns,Lappi:2003bi} implemented in the publicly available~\footnote{The \texttt{curraun} solver is available at \href{https://github.com/avramescudana/curraun/tree/wong}{github.com/avramescudana/curraun/tree/wong}.} \texttt{curraun}~solver. 

We simulate the propagation of heavy quark pairs $Q\overline Q$ that are produced back-to-back at the time $\tau_\mathrm{form}=1/(2m_T)$ where $m_T=m^2+p_T^2(\tau_\mathrm{form})$. 
The initial heavy quark transverse momentum spectrum is obtained from FONLL~\cite{Cacciari:1998it,Cacciari:2001td} that implements a pQCD calculation including next-to-leading logarithmic corrections.  
The initial rapidity is set to $\eta^Q(\tau_\mathrm{form})=\eta^{\overline Q}(\tau_\mathrm{form})=0$ in our boost-invariant setup. The initial quark colors are random and independent of each other, corresponding to the color structure  in the dominant production mechanism $g+g\to Q+\overline Q$. From these simulations we extract the heavy quark distributions as a function of proper time since the pair formation: $\Delta \tau = \tau - \tau_\mathrm{form}$. This evolution is illustrated in Fig.~\ref{fig:sketch_quark_pair}.

\textit{Glasma signals in heavy quark production---}Numerical simulations of the heavy quark pair evolution in the evolving glasma background allow us to identify observables that are sensitive to glasma dynamics in heavy ion collisions.
The strength of the glasma fields is controlled by the saturation scale of the nucleus, for which we use  $Q_s=2\,\mathrm{GeV}$ as a realistic value for central heavy ion collisions at LHC energies.
We focus on charm quark production, as the charm is heavy enough to ensure that the quarks are produced early in the collision but also not too heavy to keep production rates large. The case of bottom quark production and dependence on $Q_s$ is studied in the companion paper~\cite{Avramescu:2024poa}. 

We begin by considering single inclusive charm production as a function of the charm quark transverse momentum $p_T$. There are two separate effects at the initial state (before the QGP formation) that can modify the heavy quark spectrum. First, the produced charm quarks are deflected by the strong  color fields in the glasma phase. This is the effect we get from our numerical simulations described above. Additionally, the partonic structure of the nucleon bound within a heavy nucleus differs from a free proton or neutron. This effect is encoded in nuclear parton distribution functions (nPDFs)~\cite{Klasen:2023uqj}.

\begin{figure}[bt]
\includegraphics[width=0.8\columnwidth]{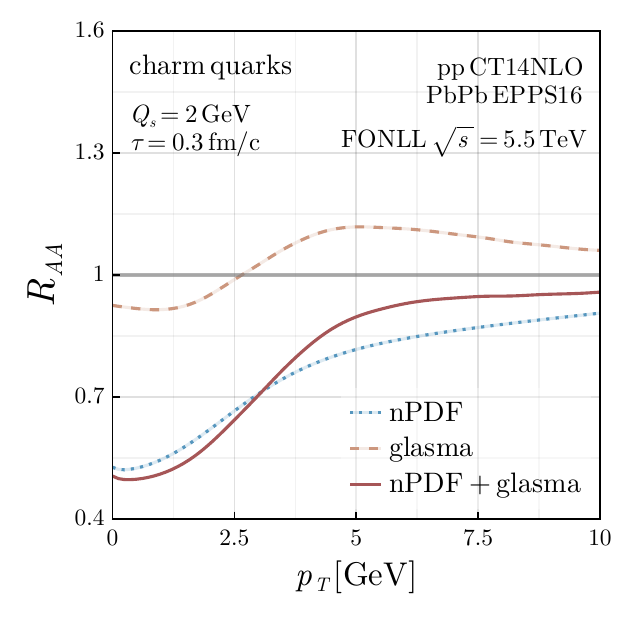}
\caption{\label{fig:raa_charm_pdf_vs_npdf}
Nuclear modification factor $R_{AA}$ for charm quarks calculated by including only the nuclear PDF effects  (\emph{nPDF}) or interactions in the glasma stage (\emph{glasma}), or both (\emph{nPDF $+$ glasma}).
} 
\end{figure}

In order to determine the relative importance of the nPDF and glasma effects on the heavy quark spectrum, we calculate the nuclear modification factor 
\begin{equation}
R_{AA} = \frac{\dd \sigma^{AA}/\dd p_T}{A^2 \dd \sigma/\dd p_T}
\end{equation} 
as a function of charm quark transverse momentum $p_T$. Note that in the absence of nuclear modification (nucleus consisting of $A$ independent nucleons) and without the modifications from the glasma  one gets $R_{AA}=1$. The obtained nuclear modification factors are shown in Fig.~\ref{fig:raa_charm_pdf_vs_npdf} at proper time $\tau=0.3\,\mathrm{fm}$ which is a typical time at which one could switch to a hydrodynamical or kinetic theory description~\cite{Kurkela:2018vqr}. In the figure, three different results are shown. First, the \emph{nPDF} curve refers to the case where no modifications from the glasma state are included, and the charm quark spectrum in the $AA$ collision differs from the $pp$ one by the fact that we use the EPPS16 nuclear parton distribution functions~\cite{Eskola:2016oht}. Note that in this case there is no time dependence in $R_{AA}$. Second, the \emph{glasma} curve only includes effects from the glasma stage, i.e.~we use free nucleon PDFs~\cite{Dulat:2015mca} also when calculating the initial $AA$ spectrum, but then follow the heavy quark evolution in the glasma stage up to $\tau=0.3\,\mathrm{fm}$. As the particle number is conserved, $R_{AA}$ must, in this case, obey a ``sum rule'' where values $R_{AA}>1$ must be compensated by $R_{AA}<1$ at another $p_T$. Finally, the \emph{nPDF+glasma} curve shows the combined effect where both the nuclear PDFs and evolution in the glasma are included. This last case corresponds to the most realistic description of the early time dynamics. The Gaussian $p_T$-broadening model developed in \cite{Avramescu:2024poa} with the width $\sigma=0.3 Q_s$ tuned to glasma simulations, largely describes the effect on $R_{AA}$.

The key finding is that the glasma stage enhances the nuclear effect obtained from nPDFs slightly, which is visible as an increased $p_T$-slope of the $R_{AA}$. This can be understood to result from the $p_T$ broadening, that pushes quarks to higher $p_T$. However, the overall effect on top of the nPDF-only result is moderate. This is expected, as previous studies including only nuclear PDF effects and heavy quark propagation in the QGP phase can successfully describe the RHIC and LHC data~\cite{Cao:2013ita,Cao:2015hia,Scardina:2017ipo,Das:2015ana,Sambataro:2022sns}, see Ref.~\cite{Cao:2018ews,Dong:2019unq} for a review. The modest effect on inclusive charm production suggests that the effect of the glasma on the initial stage is consistent with existing measurements of heavy flavor spectra.

Let us then move to heavy quark angular correlations. Such correlations can be more sensitive to early time dynamics than inclusive measurements, as the heavy quark propagation in the glasma field can easily wash out the initial back-to-back structure. 
 Furthermore, as discussed above, heavy quark pair production measurements, in practice $D\overline D$ correlations,  will be performed at the HL-LHC~\cite{ALICE:2022wwr}. 

\begin{figure*}[t]
\centering
\includegraphics[width=0.65\columnwidth]
{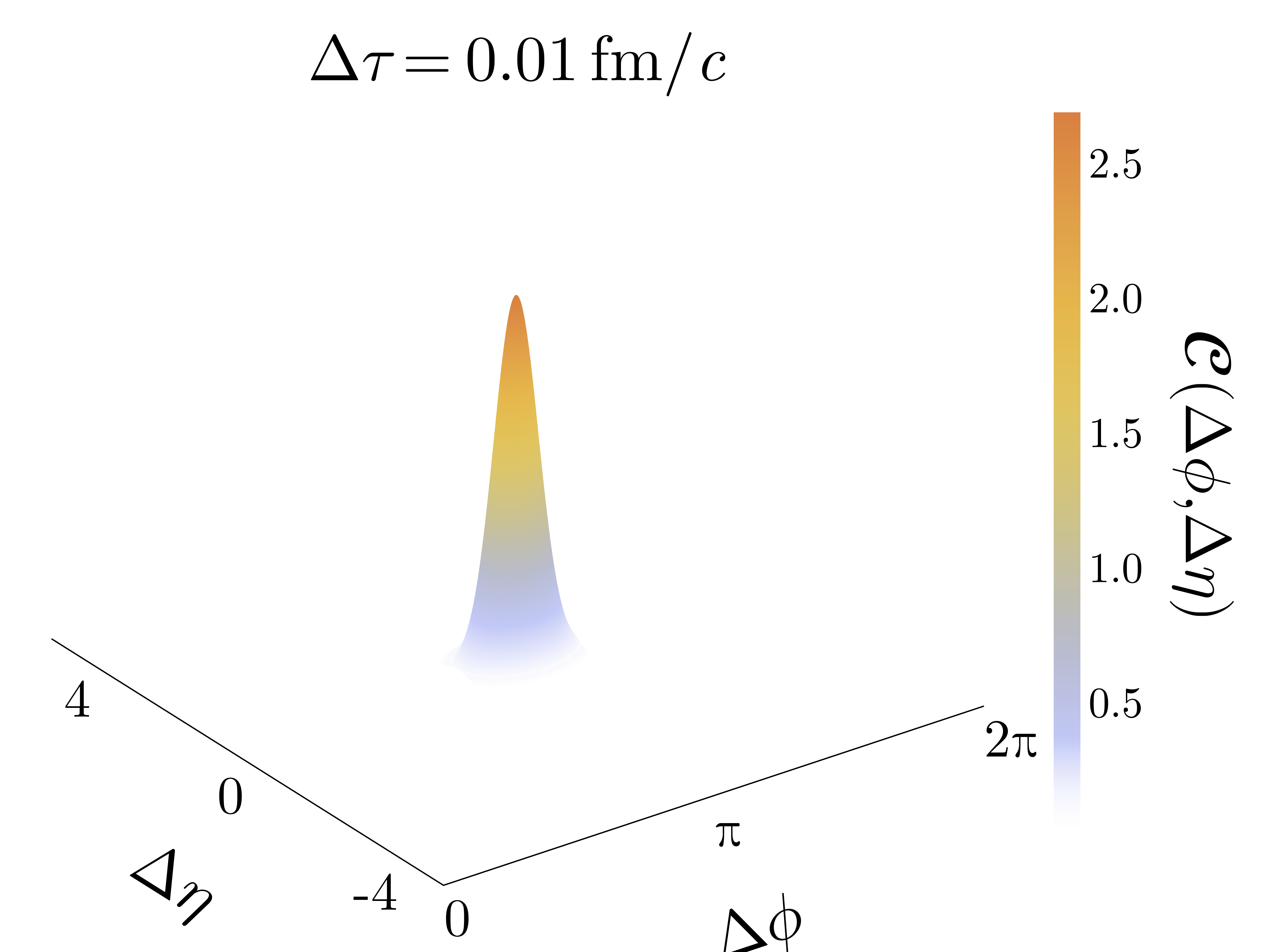}\quad
\includegraphics[width=0.65\columnwidth]
{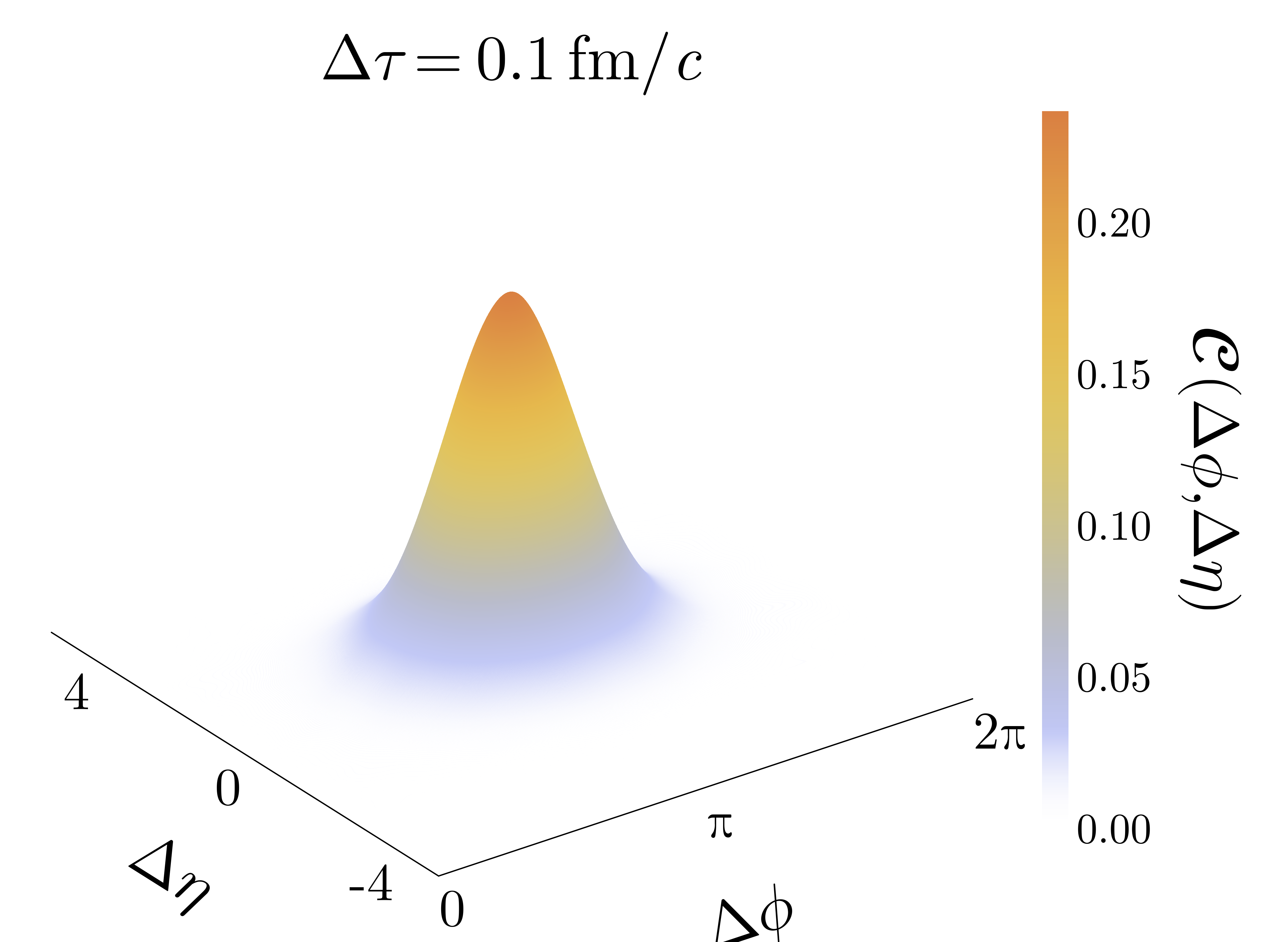}\quad
\includegraphics[width=0.65\columnwidth]
{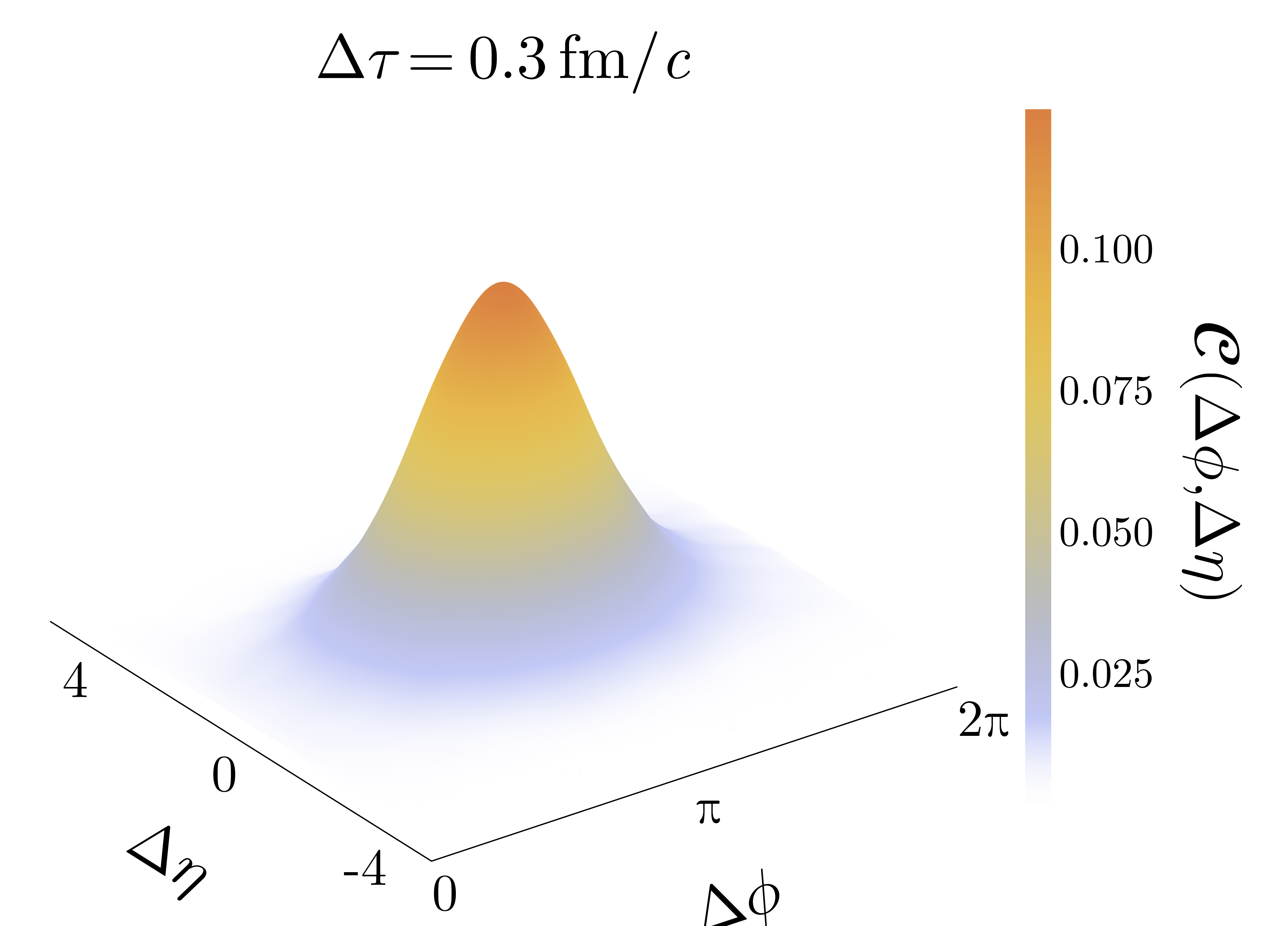}
\caption{Proper time dependence of the $c\bar c$ correlation as a function of rapidity separation $\Delta \eta$ and relative azimuthal angle $\Delta\phi$. The color illustrates the strength of the correlation function, which is normalized to unity.}
\label{fig:3dtwopartcorr}
\end{figure*}

The two-particle correlation in the glasma depends on time. As discussed above, we initialize charm quark pairs back-to-back at $\Delta\tau=0$ and extract the correlation function at later times. The evolution in the $\Delta \eta,\Delta\phi$ plane is illustrated in Fig.~\ref{fig:3dtwopartcorr} for the setup where the initial $p_T(\tau_\mathrm{form})=1\,\mathrm{GeV}$. Here the evolution is followed up to $\Delta\tau=0.3\,\mathrm{fm}/c$ in order to cover the typical timescale of the glasma stage at the LHC energies. The initial back-to-back correlation is found to be washed out quickly. In rapidity the decorrelation is visible already at very early times $\Delta\tau=0.01\,\mathrm{fm}/c$, and a similar decorrelation in the azimuthal angle becomes clearly visible at later times.
A stronger decorrelation in the longitudinal direction is consistent with the findings of Ref.~\cite{Avramescu:2023qvv}. At $\Delta\tau=0.3\,\mathrm{fm}/c$ where one typically should switch to kinetic theory followed by a hydrodynamical description of QGP~\cite{Kurkela:2018vqr}, the back-to-back peak is heavily suppressed. This suggests that a significant modification of $D\overline D$ correlations can be expected from the glasma stage. 

\begin{figure}[tb]
\includegraphics[width=0.7\columnwidth]{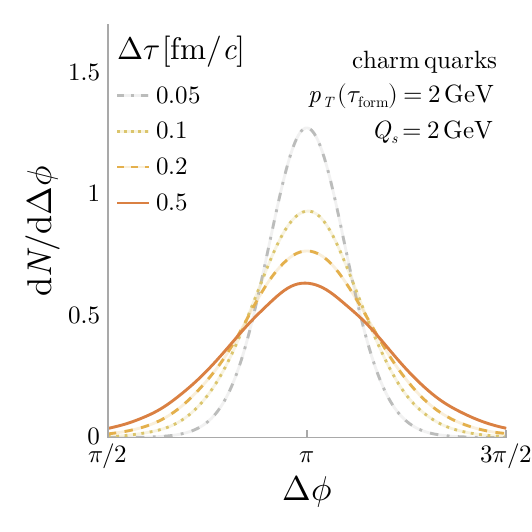}
\caption{\label{fig:dndphi_tau} Azimuthal angle correlation for charm quarks initialized with $p_T(\tau_\mathrm{form})=2\,\mathrm{GeV}$. The decorrelation is extracted at various values of proper time $\Delta\tau\in\{0.05, 0.1, 0.2, 0.5\}\,\mathrm{fm/c}$.}
\end{figure}

In order to quantify the decorrelation strength in more detail, we show the normalized two-particle correlation function for charm quark pairs in Fig.~\ref{fig:dndphi_tau} as a function of the azimuthal angle difference $\Delta\phi$ and proper time $\Delta\tau$. The initial charm quark transverse momentum is set to a typical value of $p_T(\tau_\mathrm{form})=2\,\mathrm{GeV}$. We find a rapid decorrelation away from the initial $\delta(\Delta\phi-\pi)$ peak during the first $\Delta\tau\sim 0.05\dots 0.1\,\mathrm{fm}/c$. This is the main result of our work. The decorrelation then increases with proper time, but only slowly at later times $\Delta\tau\gtrsim 0.1\,\mathrm{fm}/c$. This is because the glasma fields become weaker when the system undergoes a rapid expansion in the longitudinal direction, which is also visible as a rapidly decreasing energy density (see  Fig.~\ref{fig:sketch_quark_pair}).
Our results indicate that although the color fields become weak rather quickly in the evolution, charm quarks are produced early enough so that they have enough time to ``feel'' the strong glasma fields. This results in a significant glasma effect at an observable level, in this case in  $D\overline D$ correlations at LHC energies. 

\textit{Discussion---}In this Letter, we have demonstrated that early-time glasma fields have a large impact on heavy quark correlations in heavy ion collisions. 
In particular, the future $D\overline D$ measurements performed at the HL-LHC provide a unique opportunity to probe the pre-equilibrium glasma stage with extremely strong non-Abelian QCD color fields. Our work demonstrates that a realistic description of the glasma stage is necessary for a proper interpretation of these future correlation measurements. While this letter focuses on Pb+Pb collisions, the HL-LHC program also emphasizes O+O collisions, where non-equilibrium dynamics play a greater role in charm hadron production. O+O and pA collisions are particularly valuable for studying these effects and improving heavy flavor angular correlation measurements.

We have quantified the effect of the glasma stage on heavy quark momentum distributions by simulating the time evolution of classical test particles on the evolving glasma background. These simulations are performed using the publicly available numerical tools published in Ref.~\cite{Avramescu:2023qvv}. Based on our numerical simulations we present the first extraction of the time-dependent heavy quark pair correlations in glasma, shown in Fig.~\ref{fig:dndphi_tau}. The observed significant decorrelation in the glasma is the main result of this work. 

Furthermore, we also determine the effect of the pre-equilibrium stage on inclusive heavy quark production by combining the  effects originating from the nuclear parton distribution functions and from the glasma. Unlike in two-particle correlations, the glasma has only a moderate effect on the single inclusive heavy quark spectra, momentum broadening slightly increasing the nuclear effect quantified in terms of the nuclear modification factor $R_{AA}$ shown Fig.~\ref{fig:raa_charm_pdf_vs_npdf}. More details of the applied setup, as well as detailed analyses showing dependence on various kinematic variables and on the nuclear saturation scale, as well as additional observables, are presented in the companion paper~\cite{Avramescu:2024poa}.

Wong's equations are valid for any $Q_s/M$ but the factorization between the heavy quark production and propagation in glasma needs corrections at $Q_s/M \sim 1$. Coupling the quark production and propagation dynamically, such as solving the Dirac equation in the glasma fields \cite{Gelis:2004jp,Gelis:2005pb,Gelis:2015eua,Gelis:2019dqb,Tanji:2017xiw} addresses this but is technically challenging. Moreover, we don't consider the back-reaction of the heavy quarks on the glasma fields, valid for $M\gg Q_s$. Including the back-reaction for fast quarks $v\sim 1$ in classical Yang-Mills simulations triggers the numerical Cherenkov instability. Implicit numerical schemes suppress this only along one direction \cite{Ipp:2018hai} and are not applicable in our case.

The azimuthal decorrelation in the glasma phase  is so large, that it cannot be neglected compared to the effect of the  QGP phase, as the decorrelation width $\sigma_{\Delta\phi}$ from the glasma is comparable to that reported for $c\overline{c}$ or $b\overline{b}$ pairs in the QGP \cite{Nahrgang:2013saa}. Current QGP models, which ignore pre-equilibrium dynamics, rely on tuned parameters reflecting only the QGP phase. Including the glasma stage would reduce the QGP contribution and require parameter re-tuning. Hadronization has a sub-leading effect on angular decorrelations \cite{Cao:2014pka,Zhao:2024oma}, supporting the glasma's role as a key pre-equilibrium signal on the azimuthal decorrelations.

To follow up on this work, it would be important to combine the glasma effects  with  simulations of the subsequent stages of heavy ion collisions. In particular, our results should be coupled to a kinetic theory description of the thermalization process and to a hydrodynamically evolving QGP phase in order to obtain predictions for the $D\overline D$ correlations that can be confronted with future data. 
Our simulation can also be refined by taking into account finite nuclear geometry with event-by-event fluctuations in nucleon positions and nucleon substructure~\cite{Mantysaari:2016ykx,Mantysaari:2020axf}. This also allows us to provide predictions for small systems ($p+p$ and $p+A$) where the QGP phase should have a weaker effect on final state correlations. Additionally, higher-order effects to initial heavy quark production that modify the exact back-to-back structure can be included e.g.~using a heavy-ion event generator such as Pythia Angantyr~\cite{Bierlich:2018xfw}. Finally, the setup can be extended beyond the boost-invariant approximation employed here by performing a full 3+1D simulation as in Refs.~\cite{Schenke:2016ksl,Ipp:2017lho,Schlichting:2020wrv,Ipp:2021lwz,McDonald:2023qwc,Matsuda:2023gle,Ipp:2024ykh}. 

\begin{acknowledgments} 
\textit{Acknowledgments---}We are grateful to I.~Helenius and H.~Paukkunen for very helpful discussions about heavy quark production, the FONLL calculation, and nPDFs. D.A.~is grateful to Pol-Bernard Gossiaux for insightful discussions about $D\overline{D}$ correlations. 
This work was supported by the Research Council of Finland, the Centre of Excellence in Quark Matter (projects 346324 and 364191), and projects 338263, 346567, and 359902 (H.M). D.A.~acknowledges the support of the Vilho, Yrjö and Kalle Väisälä Foundation. V. G. acknowledges support from the European Union—Next Generation EU through the program
PRIN2022 (Cod. 2022SM5YAS). D.M.~acknowledges support from the Austrian Science Fund (FWF) projects P~34764 (Grant DOI 10.55776) and P~34455 (Grant DOI 10.55776). This work was also supported under the European Union’s Horizon 2020 research and innovation programme by the European Research Council (ERC, Grant Agreements  No. ERC-2023-101123801 GlueSatLight and ERC-2018-ADG-835105 YoctoLHC) and by the STRONG-2020 project (Grant Agreement No. 824093). The content of this article does not reflect the official opinion of the European Union and responsibility for the information and views expressed therein lies entirely with the authors. 
Computing resources from CSC – IT Center for Science in Espoo, Finland and the Finnish Grid and Cloud Infrastructure (persistent identifier \texttt{urn:nbn:fi:research-infras-2016072533}) were used in this work.
\end{acknowledgments}

\bibliographystyle{JHEP.bst}
\bibliography{refs}

\end{document}